\documentclass[aps,prl,superscriptaddress,twocolumn,showpacs]{revtex4}
\usepackage{amsmath}
\usepackage{graphicx,epsfig}

\newcommand{\be}{\begin{equation}}
\newcommand{\ee}{\end{equation}}
\newcommand{\nn}{\nonumber\\}

\newcommand{\ba}{\begin{eqnarray}}
\newcommand{\ea}{\end{eqnarray}}

\newcommand{\bpm}{\begin{pmatrix}}
\newcommand{\epm}{\end{pmatrix}}

\newcommand{\abar}{a^+}

\newcommand{\gammabar}{\gamma^+}

\newcommand{\psibar}{\psi^+}

\begin{document}

\title{Hierarchy of Spin and Valley Symmetry Breaking in \\
Quantum Hall Single Layer Graphene}
\author{Zhihua Yang}
\affiliation{Department of Physics, Sungkyunkwan University, Suwon
440-746, Korea}
\author{Jung Hoon Han}
\email[Electronic address:$~~$]{hanjh@skku.edu}
\affiliation{Department of Physics, Sungkyunkwan University, Suwon
440-746, Korea}

\begin{abstract} We explore several microscopic mechanisms for
breaking the $n=0$ fourfold Landau level degeneracy in a single-layer
graphene. Valley-scattering random potential, Zeeman interaction, and
electron-phonon coupling are considered in the presence of
SU(4)-symmetric Coulomb exchange interaction. Among all the
mechanisms considered, it is the electron-phonon coupling combined
with the Zeeman interaction which leads to the full splitting of the
$n=0$ Landau levels. A recent controversy of ``valley-first" or
``spin-first" breaking of SU(4) symmetry of the $n=0$ graphene Landau
level is examined in light of our results. Existence of midgap states
between Landau levels of opposite valley polarity are demonstrated.
\end{abstract}

\maketitle

Spin and valley degeneracy of a single-layer graphene sheet subject
to a perpendicular magnetic field gives rise to a quantized Hall
conductance\cite{graphene-QHE-geim,graphene-QHE-kim} $\sigma_{xy}$
that changes in multiples of four units of conductance quantum
$e^2/h$: $\sigma_{xy} = 4 (e^2/h) (n+1/2)$, $n$=integer. The
conductance changes from (taking $e^2/h \equiv 1$) $-2$ to $+2$ as
the chemical potential passes through the central, $n=0$ Landau
level (LL) which is fourfold degenerate. When the strength of the
magnetic field increases, these fourfold degenerate LL's split in
energy into two
sublevels\cite{kim-high-field06,geim-high-field07,kim-high-field07},
and accordingly the Hall conductance steps occur at $-2, 0$, and
$+2$. At even higher fields the fourfold degeneracy is broken
completely, leading to the sequence $\sigma_{xy} =
-2,-1,0,1,2$\cite{kim-high-field06,geim-high-field07,kim-high-field07}.

The manner of the breaking of fourfold degeneracy of the central LL
has been discussed in a number of
papers\cite{nomura,ALL,goerbig,alicea,herbut,ezawa,lederer,sheng,gusynin}.
The SU(4) symmetry of the pristine $n=0$ graphene LL is broken down
either spontaneously from interaction effects, or due to various weak
symmetry-breaking terms such as lattice effects, Zeeman splitting,
etc. Since the degeneracy arises from SU(2) symmetry of the electron
spin and another SU(2) symmetry of the valley, the main focus of
discussion has been whether the spin-symmetry or the valley-symmetry
breaking occurs first. Abanin, Levitov, and Lee argued that the spin
symmetry breaking  should occur first, turning the edge of the Hall
sample into conducting channels\cite{ALL}. An experiment carried out
shortly thereafter seems to confirm their
picture\cite{geim-high-field07}. On the other hand, subsequent
transport experiments\cite{ong} which found divergent longitudinal
resistance seem to rule out the existence of such gapless edge
states, and the issue of ``spin-first" or ``valley-first" symmetry
breaking appears by no means settled.

In this paper, we revisit the ``hierarchy problem" of the central LL
splitting in a single-layer graphene within the self-consistent
Hartree-Fock theory, while considering several SU(4)-symmetry
breaking terms explicitly. Following the general approach, we adopt
the continuum description of the graphene dynamics using the spinor
$\psi_{\sigma\tau } (r) =
\left(\begin{array}{c} a_{\sigma\tau} (r) \\
b_{\sigma\tau} (r) \end{array} \right)$. Spin ($\sigma =\uparrow,
\downarrow$) and valley index $\tau = \pm $ are introduced to
classify the spinors formed from $a$- and $b$-sublattice electrons.
The Landau level problem with the perpendicular magnetic field can be
treated by the Hamiltonian

\ba  H^\mathrm{K} = {\hbar \omega i} \sum_{\sigma\tau} \int d^2 r ~
\psibar_{\sigma \tau} (r)  \left(\begin{array}{cc} 0
& -a  \\
\abar & 0
\end{array} \right)\psi_{\sigma \tau} (r).  \ea
With the non-commuting operators obeying $[p_x , p_y ] = i \hbar^2 /
l_B^2$, $l_B =\sqrt{\hbar /eB}$ being the magnetic length, one can
form a set of canonical operators $a = (l_B /\sqrt{2}\hbar) (p_x
+ip_y )$, $\abar = (l_B /\sqrt{2}\hbar) (p_x -ip_y )$, $[a,
\abar]=1$. A cyclotron frequency $\omega = \sqrt{2}v_F /l_B$
($v_F$=Fermi velocity) has been introduced above.

In writing down the Hamiltonian in the manifestly SU(4)-symmetric
form above, we have implemented the rotation of the $\tau =-$ spinor,
$\psi_{\sigma -} \rightarrow \sigma_y \psi_{\sigma -}$. Using the
complete set of normalized eigenfunctions given by

\be \chi_{nm}  =  {1\over \sqrt{2}} \left( \begin{array}{c}
\mathrm{sgn}(n) \phi_{|n| -1,m} \\ i\phi_{|n|m}
\end{array} \right), ~~ \chi_{0m}  =\left( \begin{array}{c}
0 \\ \phi_{0m} \end{array} \right), \label{eq:chi-nm}\ee
one may expand the field operator as $\psi_{\sigma \tau} (r) =
\sum_{n,m} \chi_{nm\tau} (r) \gamma_{nm\sigma\tau}$. Here $\phi_{nm}$
is the oscillator wave function, $m$ is the guiding center
coordinates, and $\chi_{nm\tau}$ equals $\chi_{nm}$ defined in Eq.
(\ref{eq:chi-nm}) if $\tau=+$, but equals $\sigma_y \chi_{nm}$ when
$\tau = -$.

As the primary interest of this paper is in understanding the
mechanism of level splitting within the central LL, we carry out the
projection to $n=0$ LL states. The kinetic energy gets completely
quenched, whereas Coulomb interaction within this LL reads

\ba && H^\mathrm{C} = {1\over 2} \int_{rr'} V(r\!-\! r') \phi^*_{m_4
} (r) \phi_{m_1 } (r) \phi^*_{m_3 } (r') \phi_{m_2 } (r') \nn
&& \times \sum_{\sigma \sigma' \tau \tau'} \gammabar_{m_4 \sigma\tau}
\gammabar_{m_3 \sigma' \tau'}\gamma_{m_2 \sigma' \tau'}\gamma_{m_1
\sigma\tau} .
\ea
The reference to the LL index $n=0$ has been dropped. The summation
over the repeated guiding center coordinates is implicit. For
numerical purpose, we work with a torus geometry of dimension $L_x
\times L_y$ and use the Landau gauge for which the wave functions are

\ba \phi_{m} (r) = {1\over \pi^{1/4} L_x^{1/2}} e^{i y_m x}
e^{-{1\over 2} (y- y_m)^2 } ,  y_m = {2\pi  \over L_x} m .\ea
The Coulomb Hamiltonian in this basis reads

\ba && H^\mathrm{C} = {1\over 2}{1\over L_x L_y} \sum_{k_x, k_y}
V\left( k_x , k_y \right) e^{ -{1\over2} k_x^2 -{1\over2} k_y^2+ ik_y
(y_{m_1} - y_{m_2} + k_x )} \nn
&& \times \sum_{\sigma\sigma'\tau\tau'} \gammabar_{m_1 + m_x
\sigma\tau} \gammabar_{m_2 - m_x \sigma' \tau'}\gamma_{m_2 \sigma'
\tau'}\gamma_{m_1 \sigma\tau} , \ea
where the Fourier-transformed Coulomb potential $V(k) = \int d^2 r
V(r) e^{ik\cdot r}$ is shown. Whereas $k_x, k_y$ run over all integer
multiples of $2\pi/L_x$ and $2\pi/L_y$, the guiding center
coordinates $m_1$ and $m_2$ span 1 through $N_\phi$, the number of
flux through the lattice $N_\phi$ given by $2\pi N_\phi = L_x L_y$.

Having established a discretized Hamiltonian, we solve it within the
Hartree-Fock theory using the self-consistent parameter

\ba \Delta_{\sigma_1 \tau_1, \sigma_2 \tau_2} (m_1, m_2 )= \langle
\gamma^+_{m_1 \sigma_1 \tau_1} \gamma_{m_2 \sigma_2 \tau_2} \rangle
\label{eq:SC-parameter}\ea
with an arbitrary pair of guiding center indices $m_1, m_2$ and the
spin-valley indices. For the reason that Hartree term offers only a
chemical potential shift for the uniform solutions we find, and that
Hartree interaction does not break the SU(4) symmetry, we will be
exclusively concerned with the exchange Hamiltonian, $H^\mathrm{EX}$.

Among the possible SU(4) symmetry-breaking terms we consider the
following three: (i) Zeeman field: $H^\mathrm{B}=B_\sigma
\sum_{m\sigma\tau}\sigma \gamma_{m\sigma\tau}^+
\gamma_{m\sigma\tau}$, (ii) Valley-scattering impurity: $
H^\mathrm{imp}=\sum_{m\sigma\tau}V_m \tau\gamma_{m\sigma\tau}^+
\gamma_{m\sigma\tau}$. We take $V_m$ as a random number of width $W$:
$V_m\in [-W/2, W/2]$, and (iii) Valley-scattering electron-phonon
coupling\cite{nomura-ryu-lee}: $H^\mathrm{el-ph}=-U\int_{r}
\sum_{\sigma \sigma' \tau}
(\psi_{\sigma\tau}^\dag\sigma_x\psi_{\sigma\bar{\tau}})
(\psi_{\sigma'\bar{\tau}}^\dag\sigma_x \psi_{\sigma'\tau})$
($\overline{\tau} = -\tau$). Projected onto the central LL and
treated in the mean-field manner, this last Hamiltonian becomes

\ba H^\mathrm{el-ph}_\mathrm{MF}=-U'\sum_{m\sigma\tau}
\Bigl(\sum_{\sigma'}\Delta_{\sigma' \tau,\sigma'
\bar{\tau}}(m,m)\Bigr)\gamma_{m\sigma \bar{\tau}}^+ \gamma_{m\sigma
\tau}, \label{eq:HUMF} \ea
where $U'=U/\sqrt{2\pi}$\cite{comment}. For convenience, we use $U$
instead of $U'$ from now.

The total Hamiltonian we will consider is
$H=H^\mathrm{EX}+H^\mathrm{B}+H^\mathrm{imp}+H^\mathrm{el-ph}_\mathrm{MF}$.
Although all these terms individually may have been analyzed in
various ways in the past, it is our belief that their combined
effects and possible competition among different symmetry-breaking
tendencies have never been studied in the presence of the Coulomb
exchange interaction in a self-consistent manner. We made extensive
numerical simulation at zero temperature to identify which of the
combinations of the Zeeman, impurity, and electron-phonon
coupling-induced interactions would lead to the full splitting of the
fourfold degeneracy. Both quarter-filled and half-filled cases were
examined. The 3/4-filled case can be deduced by symmetry from the
results of 1/4-filled case.

With $H^\mathrm{EX}$ alone and at half-filling, the initial fourfold
degeneracy of the LL is split into two sublevels with energies at
$\pm E^\mathrm{EX}$, where the scale $E^\mathrm{EX}$ is set by the
exchange energy interaction. In our convention, $e^2 /\kappa l$
($\kappa$=dielectric constant) is taken to unity, and in such a unit
we obtain $E^\mathrm{EX}\approx 0.5$. For the quarter-filled case, a
similar situation arises with one LL at an energy below the chemical
potential and three degenerate LL's whose energy lies above it. From
these exercises we learn that the full energy splitting of the
central LL requires more than the Coulomb exchange effect alone.
Which of the spin and valley symmetry remains intact is completely
arbitrary at this point.

\begin{figure}[t]
\includegraphics[width=0.8\columnwidth,bb=17 30 312 224]
{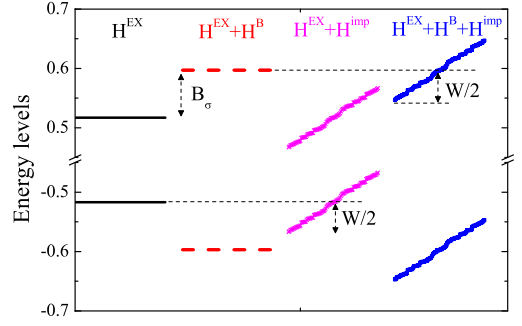} \caption{(color online) Plot of energy levels at
half-filling obtained from self-consistent solutions of
$H^\mathrm{EX}$ (black solid lines),  $H^\mathrm{EX} + H^\mathrm{B}$
(red dash lines), $H^\mathrm{EX} + H^\mathrm{imp}$ (magenta
crosses), and $H^\mathrm{EX} + H^\mathrm{B} + H^\mathrm{imp}$ (blue
filled squares). The impurity broadening is $W = 0.1$ and the Zeeman
field is $B_\sigma=0.08$. The system size used is $N_\phi = 50$. }
\label{fig:HexHBHimp}
\end{figure}

Still at half-filling, inclusion of the Zeeman field to the Coulomb
exchange now ensures that the symmetry breaking occurs along the spin
direction, with the LL energies at $\pm (E^\mathrm{EX}+B_\sigma )$.
The valley-SU(2) symmetry is preserved under the addition of
$H^\mathrm{B}$. It thus appears that more than one SU(4) symmetry
breaking mechanism need to be present to fully split the degeneracy.
We find that further inclusion of the valley-scattering impurity,
$H^\mathrm{EX}+ H^\mathrm{B} + H^\mathrm{imp}$, does not result in
additional splitting of the levels. The previously twofold degenerate
states for each guiding center $m$ undergoes splitting by $\pm V_m$,
and give rise to broadened energy levels of width $W$. The
numerically obtained energy levels for several combinations of terms
at half-filling can be found in Fig. \ref{fig:HexHBHimp}.

Actually, the Coulomb exchange Hamiltonian perturbed by two kinds of
Zeeman fields separately acting on the spin and the valley spaces, as
in $H^\mathrm{EX}+ B_\sigma \sum_{m\sigma\tau} \sigma
\gamma_{m\sigma\tau}^+ \gamma_{m\sigma\tau}+ B_\tau
\sum_{m\sigma\tau} \tau \gamma_{m\sigma\tau}^+ \gamma_{m\sigma\tau}$,
does exhibit a full lifting of the fourfold degeneracy with the
energies given at $( E^\mathrm{EX}\!+\!B_M ) \!\pm\! B_m$ and at
$-(E^\mathrm{EX}\!+\! B_M ) \!\pm\! B_m$. Here $B_M$ and $B_m$ refer
to the larger and the smaller of the two Zeeman fields, respectively.
One can also see that the electron-phonon coupling
$H^\mathrm{el-ph}_\mathrm{MF}$ provides the required valley Zeeman
field, acting along the $x$-axis of the valley spin.

\begin{figure}[ht]
\includegraphics[width=1.0\columnwidth,bb=14 9 380 222]
{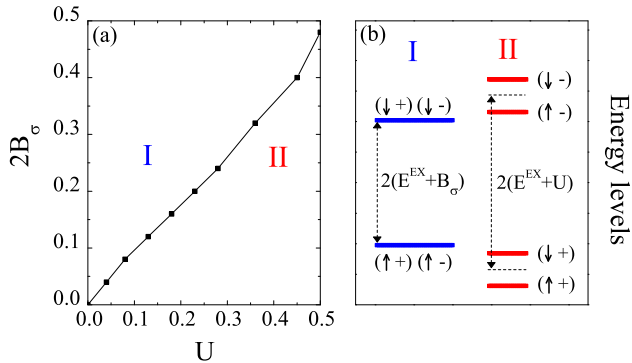} \caption{(color online) (a) Phase diagram of
the central LL at half-filling with varying Zeeman field
($2B_\sigma$) and electron-phonon coupling strength ($U$). In region
I, only the spin degeneracy is broken. In region II, both spin and
valley symmetries are lost, but the main level splitting takes place
in the valley direction. A first-order phase boundary separates the
two regions. (b) Schematic energy levels in each region. The energy
level separations and spin-valley quantum numbers for each LL are
specified.} \label{fig:half-filling-phase}
\end{figure}

Aided by these ideas, we next consider $H^\mathrm{EX} + H^\mathrm{B}
+ H^\mathrm{el-ph}_\mathrm{MF}$. Figure \ref{fig:half-filling-phase}
shows the phase diagram for such a model, at half-filling, spanned by
two interaction parameters $(U, 2B_\sigma)$. There are two phases
found here, called I and II, distinguished by the number of levels
split. In region I, where the Zeeman effect dominates, only one,
spin-polarized level splitting is observed. The valley-splitting
order parameter $\Delta_{\sigma-,\sigma+}(m,m)$ becomes zero in this
region, still preserving the valley symmetry. In the $U$-dominated
region II, the main polarization direction is along the valley axis,
and the Zeeman field contributes to the sublevel splitting equal to
$2B_\sigma$. Here indeed, the full breaking of the fourfold
degeneracy is obtained. The phase boundary taking place along
$2B_\sigma \approx U$ in the phase diagram is first-order.

\begin{figure}[ht]
\includegraphics[width=1.0\columnwidth,bb=14 9 380 222]
{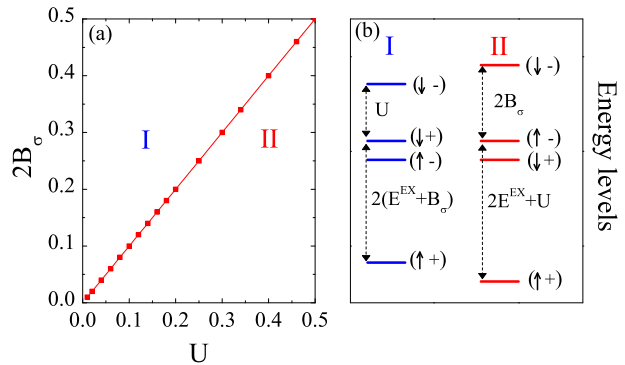} \caption{(color online) Phase diagram of
the central LL at quarter-filling. Meaning of the symbols are the
same as in Fig. \ref{fig:half-filling-phase}. Only one LL lies below
the Fermi level here. } \label{fig:quarter-filling-phase}
\end{figure}

The phase diagram for quarter-filling is similar, as shown in Fig.
\ref{fig:quarter-filling-phase}. The phase boundary now taking place
exactly at $2B_\sigma =U$ separates the ``spin-first"-split region I
from the ``valley-first"-split region II. The two centrally located
LL's cross in energy at the phase boundary. For instance,
$E(\uparrow, -) < E(\downarrow, +)$ energy hierarchy in region I
crosses over to $E(\uparrow, -) > E(\downarrow, +)$ in region II. We
have checked that the inclusion of the impurity does not alter the
basic features of the phase diagram shown in Figs.
\ref{fig:half-filling-phase} and \ref{fig:quarter-filling-phase} as
long as $W$ remains small compared to both $U$ and $2B_\sigma$.

The relevant energy scales $2B_\sigma$ and $U$ in a graphene layer
are comparable, as recently discussed in Ref. \cite{nomura-ryu-lee}.
Both energy scales are of order $B_\sigma$ in units of
[Kelvin/Tesla], and therefore it should be quite possible that
graphene samples with either $2B_\sigma > U$ or $2B_\sigma < U$
exist. Then according to Ref. \cite{ALL}, the edge of the half-filled
graphene quantum Hall system can be either conducting ($U<2B_\sigma$)
or insulating ($U>2B_\sigma$). Another interesting possibility
suggested by our search is the transition between the two LL
splitting scenarios driven by the relative strengths of
electron-phonon coupling and Zeeman energies. We speculate that
applying a mechanical pressure, such as stretching, to the graphene
will influence $U$ without changing the Zeeman energy, and might
allow one to probe the phase transition between regions I and II. The
bond-CDW order associated with region II at half-filling and regions
I and II for quarter-filling should leave a mark in the electronic
spectrum, which can be probed by STM.

A Landau level with a quantum number $(\sigma, \tau)$ occurs at the
same energy as another state with $(\sigma, \overline{\tau})$,
whereas the same is not true with $(\overline{\sigma}, \tau)$ due to
the Zeeman splitting. One can then imagine domain walls separating
the two LL states with opposite valley polarities in a macroscopic
sample. Assuming the spinless case for simplicity, the physics of
such a domain wall can be captured in a set of differential equations

\ba i (\partial_y \!+\! y_k ) [u_- \!+\! v_+] \!+\! m_\mathrm{v}(y) [
u_+ \!-\! v_- ] &=& -\varepsilon_k [u_+ \!-\! v_- ] \nn
i (\partial_y \!-\! y_k ) [u_+ \!-\! v_-] \!-\! m_\mathrm{v}(y) [ u_-
\!+\! v_+ ] &=& -\varepsilon_k [u_- \!+\! v_+ ]\nn
i (\partial_y \!+\! y_k ) [u_- \!-\! v_+] \!+\! m_\mathrm{v}(y) [ u_+
\!+\! v_- ] &=& \varepsilon_k [u_+ \!+\! v_- ]\nn
i (\partial_y \!-\! y_k ) [u_+\!+\! v_-] \!-\! m_\mathrm{v}(y) [ u_-
\!-\! v_+ ] &=& \varepsilon_k [u_- \!-\! v_+ ] .
\label{eq:diff-eq-with-mass}\ea
We have written the eigenfunction associated with the $a$- and
$b$-sublattice as $(u_\tau ,v_\tau )$, also distinguished by their
valley index $\tau$, used the linear gauge ($A_x = - B y, A_y = 0$),
and taken out the $x$-dependence of the wave function as $(u_\tau,
v_\tau) \rightarrow e^{ik x} (u_\tau (y), v_\tau (y))$. The
$y$-dependent mass gap due to the inter-valley scattering is written
$m_\mathrm{v}(y)$, and $y_k$ abbreviates $y\!-\!k$.

\begin{figure}[t]
\includegraphics[width=0.65\columnwidth,bb=32 18 449 314]    
{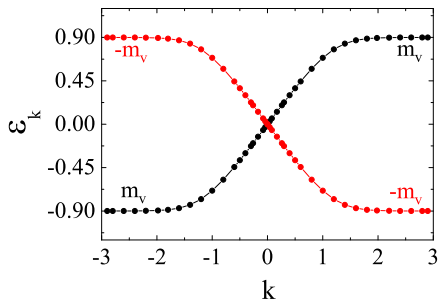} \caption{(color online) The energy
$\varepsilon_k$ obtained from Eq. (\ref{eq:energy-eigenvalues}) with
$m_\mathrm{v}=0.9$.} \label{fig:energy-crossing}
\end{figure}

As an example of the influence of the sign change of the mass gap on
the energy spectra, consider the case with an abrupt sign change as
in $m_\mathrm{v} (y) = m_\mathrm{v} \mathrm{sgn} (y)$. Away from
$y=0$ the mass gap is uniform, and a pair of solutions with
$\varepsilon_k = \pm m_\mathrm{v}$ is found for $u_+ = v_- = 0$, and
$u_- = \pm v_+ = \phi (y -k)$ respectively, where $\phi (y-k)$ is
the Gaussian function peaked at $y=k$. Since $k$ is still a good
quantum number, the eigen energies $\varepsilon_k$ can be solved for
each $k$ separately. By matching the wave functions at $y=0$ we
derive the following equation determining the energies,

\ba {m_\mathrm{v} \! - \! \varepsilon_k  \over m_\mathrm{v} \!+\!
\varepsilon_k} = {H(p_k , k) H'(p_k , -k) \over H(p_k , -k)H'(p_k ,
k) }, \label{eq:energy-eigenvalues}\ea
where $p_k =(\varepsilon_k^2 \! - \! m_\mathrm{v}^2) /2 \!<\! 0 $,
and $H(p_k ,k)$ is Hermite polynomial of negative order $p_k$.
Self-consistently solving the equation for each $k$ gives rise to the
energy band shown in Fig. \ref{fig:energy-crossing}.

The level crossing predicted here will be of particular relevance if
the chemical potential should lie between the two valley-split LL's.
In such a case, a pair of gapless one-dimensional channels will cross
the Fermi level, similar to the edge channel in the spin-split case
first discussed in Ref. \cite{ALL}. Unlike the edge channels, the
metallic channel predicted here can be formed at the bulk whenever a
domain boundary separates the opposite valley states. Referring to
our phase diagram in Fig. \ref{fig:half-filling-phase}, the two LL's
lying closest to the Fermi level always carry opposite spins,
therefore a domain wall state connecting the occupied and the
unoccupied LL's will have to involve twist in both spin and valley.
On the other hand, the quarter-filled case (region I) offer a better
chance for observing the domain wall between two spin-polarized LL's
which differ only in their valley polarities. We therefore suggest
that the ``spin-first" split LL with $\sigma_{xy} = -1$ is the most
likely platform to observe the metallic domain walls.

In summary, we explored the hierarchy issue of the central LL
symmetry breaking of a graphene layer. Self-consistent Hartree-Fock
theory was employed, taking into consideration several kinds of
SU(4) symmetry breaking terms. The competitive nature of the
valley-splitting (due to electron-phonon interaction) and
spin-splitting (due to Zeeman interaction) leads to a phase diagram
with either ``spin-first" or ``valley-first" level splitting.
Existence of a new kind of gapless state when LLs with opposite
valley polarities form a domain wall is demonstrated.

\acknowledgments  H. J. H. is supported by the Korea Science and
Engineering Foundation (KOSEF) grant funded by the Korea government
(MEST) (No. R01-2008-000-20586-0), and in part by the Asia Pacific
Center for Theoretical Physics. One of us (H. J. H.) wishes to thank
P. Kim for hospitality during the completion of this work and for
discussion.


\begin{thebibliography}{99}
\bibitem{comment} A more general valley symmetry-breaking order
parameter $\Delta_{\sigma\tau, \sigma \overline{\tau}} (m, m')$ can
be chosen with $m \neq m'$. Perhaps due to the localized nature of LL
wave functions, actual self-consistent values are negligibly small
unless $m = m'$.
\end{thebibliography}
\end{document}